\documentclass[fleqn,usenatbib]{mnras}

\usepackage{mathptmx}
\usepackage[T1]{fontenc}
\usepackage{ae,aecompl}
\usepackage{lineno}
\usepackage{graphicx}	% Including figure files
\usepackage{amsmath}	% Advanced maths commands
\usepackage{amssymb}	% Extra maths symbols
\usepackage{CJKutf8}

\newcommand{\nustar}{\textit{NuSTAR }}
\newcommand{\suzaku}{\textit{Suzaku }}

\newcommand{\xmm}{{\it XMM-Newton }}

\title[Spectral Fitting of WKK~4438]{The Ultra-Fast Outflow of WKK~4438: \suzaku and \nustar X-ray Spectral Analysis}

\author[J. Jiang et al.]{Jiachen Jiang\begin{CJK*}{UTF8}{gbsn}
(姜嘉陈)
\end{CJK*}$^{1}$\thanks{E-mail: jj447@cam.ac.uk}, Dominic J. Walton$^{1}$, Michael L. Parker$^{2}$, 
\newauthor and Andrew C. Fabian$^{1}$\\
\\
$^{1}$Institute of Astronomy, Univeristy of Cambridge, Madingley Road, CB3 0HA Cambridge, UK\\
$^{2}$European Space Agency (ESA), European Space Astronomy Centre (ESAC), E-28691 Villanueva de la Ca\~nada, Spain\\
}
\date{Accepted XXX. Received YYY; in original form ZZZ}

\pubyear{2018}

\begin{document}
\label{firstpage}
\pagerange{\pageref{firstpage}--\pageref{lastpage}}
\maketitle

\begin{abstract}
Previous X-ray spectral analysis has revealed an increasing number of AGNs with high accretion rates where an outflow with a mildly relativistic velocity originates from the inner  accretion disk. Here we report the detection of a new ultra-fast outflow (UFO) with a velocity of $v_{\rm out}=0.319^{+0.005}_{-0.008}c$ in addition to a relativistic disk reflection component in a poorly studied NLS1 WKK~4438, based on archival \nustar and \suzaku observations. The spectra of both \suzaku and \nustar observations show an Fe~\textsc{xxvi} absorption feature and the \suzaku data also show evidence for an Ar~\textsc{xviii} with the same blueshift. A super-solar argon abundance ($Z^{\prime}_{\rm Ar}>6Z_{\odot}$) and a slight iron over-abundance ($Z^{\prime}_{\rm Fe}=2.6^{+1.9}_{-2.0}Z_{\odot}$) are found in our spectral modelling. Based on Monte-Carlo simulations, the detection of the UFO is estimated to be around at 3$\sigma$ significance. The fast wind most likely arises from a radius of $\geq20r_g$ away from the central black hole. The disk is accreting at a high Eddington ratio ($L_{\rm bol}=0.4-0.7L_{\rm Edd}$). The mass outflow rate of the UFO is comparable with the disk mass inflow rate ($\dot M_{\rm out}>30\%\dot M_{\rm in}$), assuming a maximum covering factor. The kinetic power of the wind might not be high enough to have influence in AGN feedback ($\dot E_{\rm wind}/L_{\rm bol}\approx 3-5\%$) due to a relatively small column density ($12^{+9}_{-4}\times10^{22}$~cm$^{-2}$). However note that both the inferred velocity and the column density could be lower limits owing to the low viewing angle ($i=23^{+3}_{-2}$$^{\circ}$).
\end{abstract}


%%%%%%%%%%%%%%%%%%%%%%%%%%%%%%%%%%%%%%%%%%%%%%%%%%

%%%%%%%%%%%%%%%%% BODY OF PAPER %%%%%%%%%%%%%%%%%%

\section{Introduction} \label{introduction}

Recently there has been increasing evidence showing the presence of absorption line features above 7~keV in the X-ray band of various sources, including Active Galactic Nuclei (AGNs) \citep[e.g.][]{chartas02, chartas03,cappi06,tombesi10,tombesi14,tombesi16}. These absorption line features are commonly interpreted as blueshifted Fe \textsc{xxv} or Fe \textsc{xxvi} K absorption in a highly ionized environment ($\log(\xi/$erg\,cm\,s$^{-1}$)>3) and can occasionally correspond to a very large line-of-sight outflow velocity of up to $0.2-0.4c$ \citep[e.g.][]{tombesi10}. Such outflows are often referred to as ultra-fast outflows (UFOs), and lie in the mildly relativistic regime indicating that the outflow is driven from the inner accretion disk. 

There are two key technical challenges when it comes to searching for ultra-fast outflows. The first is that they are close to the upper limit of the instrumental effective area of most soft energy cameras, such as \suzaku XIS and \xmm EPIC. Elements lighter than iron are generally fully ionized and therefore show weak or absent absorption features in the soft energy band. Only strong iron absorption feature remains above 7~keV where the signal-to-noise (S/N) and the spectral resolution are worse than in the soft band. The second is that the broadband continuum needs to be correctly modeled in order robustly determine the key absorption parameters \citep[e.g.][]{zoghbi15}. This is particularly critical for AGN sources, which often exhibit complex X-ray spectra with strong reflection.

One popular theory for the origin of these extreme outflows is that the radiation pressure due to a high accretion rate drives the UFO (e.g. PDS~456; \citealt[][]{matzeu17}). It is therefore interesting to note the discovery of UFOs in ultraluminous X-ray sources \citep[ULXs; e.g.][]{pinto16, walton16, kosec18}, which appear to be sources accreting above their Eddington limit. Another ideal population to test this theory are Narrow Line Seyfert 1 (NLS1) galaxies. NLS1s are characterized by having low-mass, high-accretion-rate black holes in the center. For example, IRAS~13224$-$3809 accretes around the Eddington limit and shows a flux-dependent blueshifted Fe absorption feature above 8~keV \citep[e.g.][]{parker17a, parker17b,pinto17tmp,jiang18_tmp} which is interpreted as a UFO with velocity up to $0.236\pm0.006c$. 

WKK~4438 is a nearby (z=0.016) NLS1 galaxy hosting a low-mass supermassive black hole $M_{\rm BH}=2\times10^{6}M_{\odot}$ \citep[measured by the H$\beta$ line width in the optical band,][]{malizia08}. In this work, we analyse its archival X-ray spectra obtained by \suzaku and \nustar satellites, which show blueshifted Fe~\textsc{xxvi} and Ar~\textsc{xviii} absorption lines in addition to a relativistic reflection component.

\section{Data Reduction} \label{data_reduction}

WKK~4438 has not been studied well in the X-ray band. The only long soft band observation is one 70~ks \suzaku observation (obsID 703011010), taken in 2012. The only archival \nustar observation (obsID 60061259002) is a 20~ks snapshot in 2013.

\subsection{\suzaku Data Reduction}

We reduced the \suzaku data with the latest HEASOFT software package. We processed the raw event files for each XIS CCD and created filtered event lists for XIS0, XIS1 and XIS3 detectors by running the \suzaku pipeline. The CALDB version we use for XIS detectors is 20160607. XSELECT was used to extract the spectral products. The size of the circular region we used to extract the source regions is 3.5$'$ in radius and the background regions were selected from the surrounding areas free from the target source and the calibration source. XISRESP was used to generate the corresponding response files with ‘medium’ resolution. The spectra and the response files of the front-illuminated CCDs (XIS0 and XIS3) were combined by using ADDSPEC. We grouped the spectra to have a minimum count of 50 per bin with GRPPHA. Hereafter, the combined spectrum of the front-illuminated CCD XIS0 and XIS3 is called FI spectrum (blue in figures) and the spectrum of the back-illuminated CCD XIS1 is called BI spectrum (green in figures). We analysed the FI and BI data over the 0.7--10~keV and 0.7--9~keV energy range, respectively. The 1.7--2.5~keV band is ignored in both spectra due to calibration issues around the Si K edge \footnote{https://heasarc.gsfc.nasa.gov/docs/suzaku/analysis/sical.html}.

\subsection{\nustar Data Reduction}

The \nustar data were reduced using the \nustar data analysis software NuSTARDAS and CALDB version 20161021. In addition to the standard science (mode 1) event files, we also extracted the products from the spacecraft science (mode 6) event files by following \citet{walton16} to maximize the exposure of the observation and thus increase the S/N of our FPM spectra. The mode 6 events contribute an additional 4~ks good exposure in addition to the mode 1 events. The total net exposure is 26~ks. The source spectra were selected from circular regions with radii of 60$''$, and the background was obtained from nearby circular regions with radii of 120$''$. Spectra were extracted from the cleaned event files using NUPRODUCTS for both FPMA and FPMB. The FPMA and FPMB spectra were grouped to have a minimum count of 50 per bin by using GRPPHA. The 3.0--50.0~keV band is considered as the background spectrum dominates above 50~keV. Two spectra were analysed independently but we only show the combined spectrum in our plots (in red) for clarity. 
%\citet{kalberla}

\section{Spectral Analysis} \label{spectral_analysis}
For the spectral analysis, we use the XSPEC(12.9.1k) software package \citep{arnaud} to fit all the spectra discussed, and $\chi^2$-statistics is considered in this work. The Galactic column density towards WKK~4438 is fixed at the nominal value $4.3\times10^{21}$\,cm$^{-2}$ \citep{willingale13} if not specified. The parameter values are reported in the source rest frame. The \texttt{tbnew} model \citep{wilms00} is used for calculating the Galactic absorption. The solar abundance is taken from \citet{wilms00}. The cross section is taken from \citet{verner96}.
\subsection{Spectral Modelling} \label{spectral_fit}

\begin{figure}
\centering
\includegraphics[width=7cm]{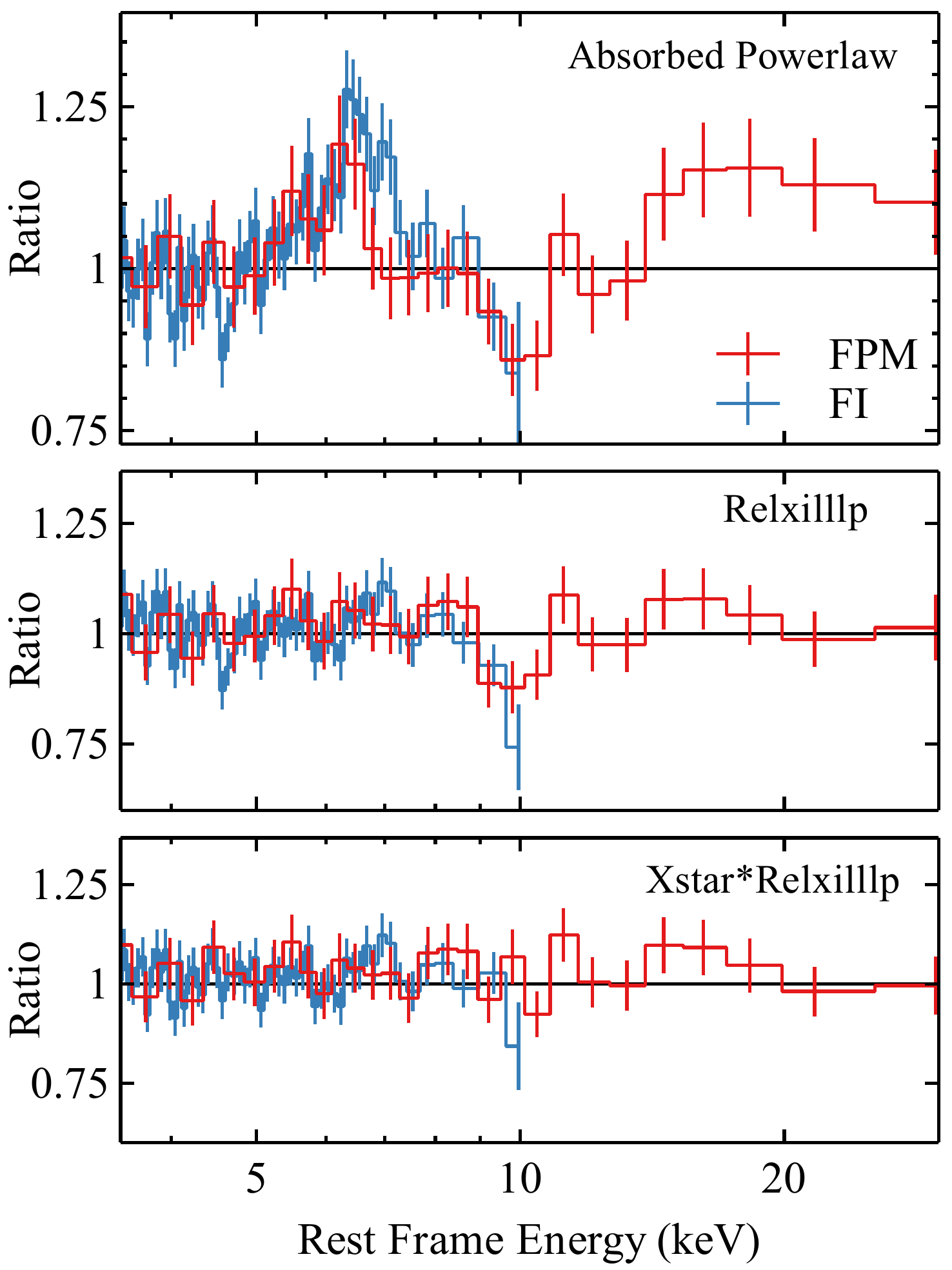}
\caption{The ratio plots against the best-fit Galactic absorbed powerlaw (top), \texttt{relxilllp} (middle) and  \texttt{xstar*relxilllp} model (bottom). Blue: \suzaku FI spectrum; red: combined FPM spectrum. \suzaku FI and BI spectra are both analysed in our work. Only FI spectrum is shown here. FPMA and FPMB are fitted independently and the combined spectrum is shown only for clarity. All the spectra show a fairly broad Fe K$\alpha$ emission line at the iron band. A Compton hump is shown above 15~keV. The \nustar spectra show a broad absorption line feature around 10~keV in the rest frame which is consistent with the \suzaku FI spectrum.}
\label{pic_fe}
\end{figure}

\begin{figure}
\centering
\includegraphics[width=7cm]{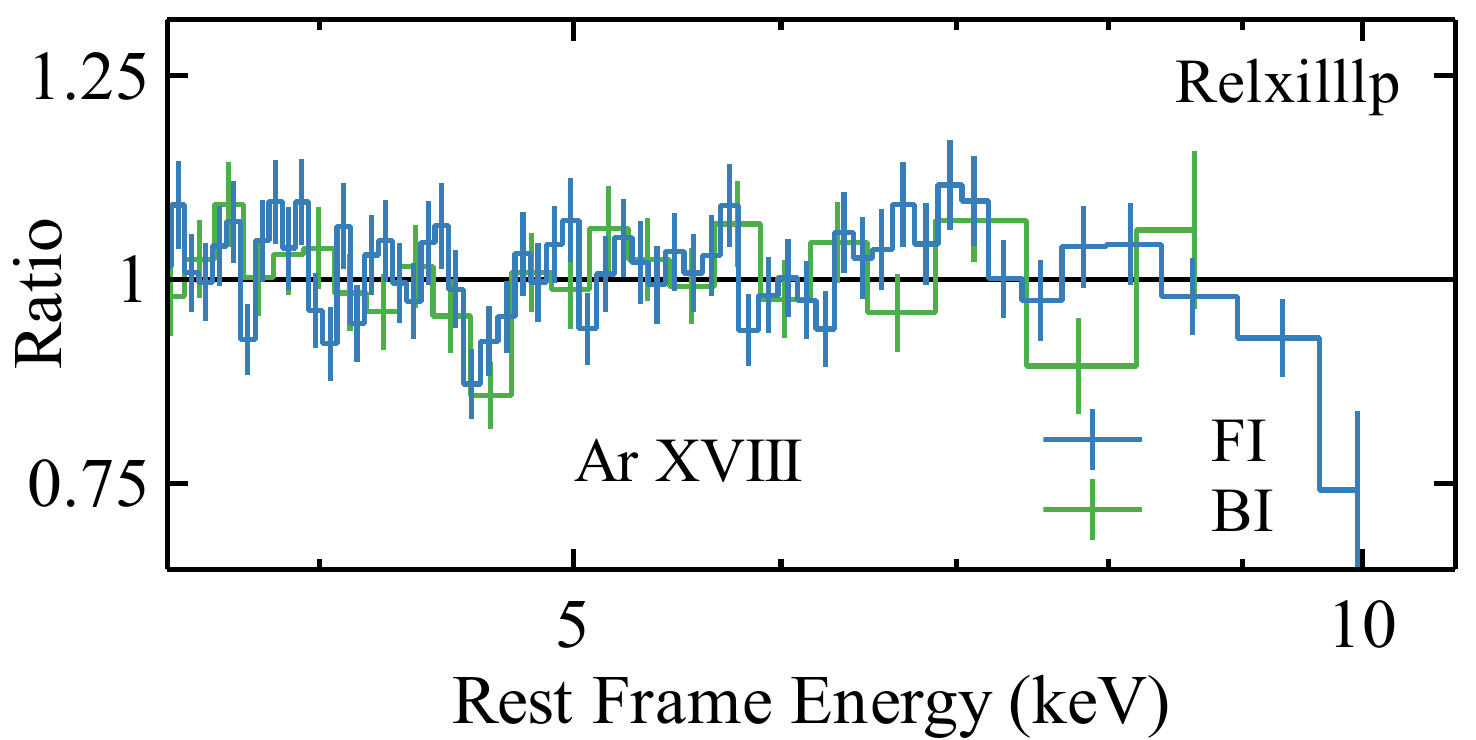}
\caption{The ratio plot against the best-fit \texttt{relxilllp} model. Blue: \suzaku FI spectrum; green: \suzaku BI spectrum. Both two \suzaku spectra show an absorption line around 4.6~keV in the rest frame, which can be explained as a blueshifted Ar~\textsc{xviii} absorption line.}
\label{pic_ar}
\end{figure}

First, we fitted the \suzaku and \nustar spectra independently with a Galactic absorbed powerlaw model. The top panel of Fig. \ref{pic_fe} shows the ratio plot against the best-fit \texttt{powerlaw} model ($\Gamma=2.00, 1.76$ for \suzaku and \nustar respectively). An Fe K$\alpha$ emission line feature in two sets of spectra and a Compton hump above 15~keV in the FPM spectra indicate a strong reflection component. By fitting the emission feature at the iron band with a simple \texttt{zgauss} line model, it improves the fit by $\Delta\chi^2=71$. The best-fit line width ($\sigma$) values are $0.41^{+0.13}_{-0.12}$~keV and $0.31^{+0.34}_{-0.16}$~keV for XIS and FPM spectra respectively, indicating a broad Fe emission line in both two sets of spectra. Both lines are broader than the instrument energy resolution. The rest frame energy of the \texttt{zgauss} line model is $6.3\pm0.1$~keV for the XIS spectra and $6.3^{+0.2}_{-0.3}$~keV for the FPM spectra. A broad absorption line feature is visible around 10~keV in the FPM spectra. The FI spectrum also shows some evidence of a broad absorption line feature at 10~keV but it lies too close to the upper limit of the observable energy range.  A second absorption line feature is visible at $\approx 4.6$~keV in both the \suzaku FI and BI spectra (see Fig. \ref{pic_ar}). As far as we know, there is no obvious calibration issue during this \suzaku observation (Yukikatsu Terada, Katja Pottschmidt priv. comm.). All three cameras of the \suzaku satellite show consistent absorption features at 4.6~keV (see Fig.\ref{pic_ar}).

Second, we fitted both the \nustar and \suzaku spectra with the relativistic disk reflection model \texttt{relxilllp} \citep{garcia14}. This model calculates consistent emissivity profile for the relativistic blurred disk iron line according to the coronal height in the lamp-post scenario. The spin parameter is fixed at its maximum value 0.998 and the inner radius of the disk is allowed to vary. The disk inner radius $R_{\rm in}$, viewing angle $i$ and the disk iron abundance $Z_{\rm Fe}$ are linked between two epochs. \texttt{relxilllp} gives a best-fit with $\chi^2/\nu = 1396.4/1438$, where $\nu$ is the number of the degrees of freedom. The relativistic blurred disk reflection model can fit the soft excess below 1~keV, the emission line at the iron band, and the Compton hump. It improves the fit by $\Delta\chi^2=533$ compared to the simple \texttt{powerlaw+zgauss} modelling. However, the broad absorption line features at 10~keV and 4.6~keV are still visible in the ratio plot (see the middle panel of Fig. \ref{pic_fe} and Fig. \ref{pic_ar} for absorption lines).

Based on the best-fit \texttt{relxilllp} model, we fitted the absorption features with two simple line models \texttt{zgauss}. The redshift parameter $z$ is fixed at the source redshift and the line energy is allowed to vary to obtain the line position in the source rest frame. The line parameters except the normalization are all linked between two observations. The inclusion of two \texttt{zgauss} models can improve the $\chi^2$ by 33 with additional 8 free parameters ($\Delta\chi^2=16$ for the $4.6$~keV absorption line and $\Delta\chi^2=17$ for the $10$~keV absorption line). %The line width of the 4.6~keV line ($4.62\pm0.02$~keV in the source rest frame) is $\sigma<0.06$~keV and that of the 10~keV absorption line ($9.8^{+0.5}_{-0.3}$~keV in the source rest frame) is $\sigma=0.3^{+0.3}_{-0.2}$~keV. Based on the best-fit line model, the equivalent widths of the 4.6~keV and 10~keV absorption lines are $24\pm8$~eV and $258^{+285}_{-184}$~eV respectively in the \suzaku data ($<41$~eV and $193^{+168}_{-130}$~eV in the \nustar data). 
The other parameters can be found in Tab.\ref{tab_line}. 

\subsubsection{Fast wind modelling} \label{fast}
In this section, we model the two absorption features with one ionized fast wind absorber.

If the absorption line at $10$~keV is interpreted as the Fe~\textsc{xxvi} Ly$\alpha$ line (6.97~keV), it corresponds to a redshift value of $z_{\rm Fe}=-0.28^{+0.03}_{-0.04}$. If the absorption line at $4.6$~keV is interpreted as the Ar~\textsc{xviii} Ly$\alpha$ line (3.32~keV), it corresponds to a redshift value of $z_{\rm Ar}=-0.281^{+0.005}_{-0.004}$. The two redshift values are consistent within the uncertainty, corresponding to a relativistic outflow with a velocity of $v\approx0.3$c.

\begin{table}
\caption{The best-fit parameters of the simple line models. The line energy is the value in the source rest frame. The numbers in the brackets show the values obtained for the \nustar spectra.}
\begin{tabular}{cccc}
\hline\hline
Energy (keV) & $\sigma$ (keV) & EW (eV) & $\Delta\chi^2$ \\
\hline
$4.62\pm0.02$ & $<0.06$  & $24\pm8$ ($<41$) & 16\\
$9.8^{+0.5}_{-0.3}$ & $0.3^{+0.3}_{-0.2}$ & $258^{+285}_{-184}$ ($193^{+168}_{-130}$) & 17 \\
\hline
\end{tabular}
\label{tab_line}
\end{table}

Finally, we modelled the absorption features in both the \nustar and \suzaku spectra with the same photoionized absorption model \texttt{xstar} \citep{xstar}. The \texttt{warmabs} model, the alternative analytical version of \texttt{xstar}, is used first to estimate the line width of the absorption lines. We fitted with the turbulent velocity as a free parameter and obtained a best-fit value of $v_{\rm tur}>6000$\,km\,s$^{-1}$ at the 2$\sigma$ confidence level. Then we constructed custom absorption models with \texttt{xstar}. The grids are calculated assuming solar abundances except for that of iron and argon, a fixed turbulent velocity of $6000$\,km~s$^{-1}$ and an ionizing luminosity of $10^{43}$\,erg~s$^{-1}$. Free parameters are the ionization of the plasma ($\log\xi^{\prime}$), the column density ($N^{\prime}_{\rm H}$), the iron abundance ($Z^{\prime}_{\rm Fe}$), the argon abundance ($Z^{\prime}_{\rm Ar}$) and the redshift (z). The prime symbol is to distinguish the parameter of the outflow from that of the disk. The iron and argon abundances of the UFO ($Z^{\prime}_{\rm Ar,Fe}$) are treated as independent free parameters during our fit. The total model reads \texttt{tbnew*xstar*relxilllp} in XSPEC format.

The best-fit model parameters obtained by fitting the two sets of spectra with \texttt{tbnew*xstar*relxilllp} can be found in Table \ref{tab_fit}. The ratio plot can be found in the bottom panel of Fig.\ref{pic_fe}. One \texttt{xstar} model with a turbulent velocity of $v_{\rm tur}=6000$\,km\,s$^{-1}$ can fit two absorption lines well simultaneously and improve the fit by $\Delta\chi^2=33$ with additional 5 free parameters after stepping the redshift parameter $z$ between -0.35 and 0 with 70 steps with all the other parameters free to vary. See the left panel of Fig.\ref{pic_contour} for the constraint on the redshift parameter $z$. The STEPPAR function in XSPEC is used for the stepping purpose. The best-fit \texttt{xstar} model is shown in the bottom panel of Fig.\ref{pic_contour}.

\subsubsection{Alternative scenarios for the absorptions} \label{slow}
In this section, we test other possible solutions for the two absorption features.

First, an alternative interpretation of the 10~keV absorption feature is the Fe K edge from a slower wind. We fitted the 10~keV absorption line with a simple edge absorption model \texttt{zedge} and obtained an improved fit with $\Delta\chi^2=7.4$. The edge model provides a worse fit of the 10~keV absorption feature than the line absorption model ($\Delta\chi^2=17$, see Table \ref{tab_line}). Therefore, we exclude the edge interpretation of the 10~keV absorption feature.

Second, an alternative identification of the 4.6~keV absorption feature could be a blueshifted Ca~\textsc{xx} absorption line, with a lower blueshift value compared to the Ar~\textsc{xviii} interpretation. The rest frame energy of the dominant Ca~\textsc{xx} Ly$\alpha$1 absorption line is around 4.11~keV. %In this modelling, a second ionized absorber with a different outflowing velocity is required in addition to the fast wind absorber obtained in Section \ref{fast}. We fixed the Ar abundance $Z^{\prime}_{\rm Ar}$ of the fast wind component at the solar value and added one more \texttt{warmabs} component in XSPEC to fit the 4.6~keV absorption feature. The turbulent velocity is fixed at 100\,km~s$^{-1}$. The combination of a fast wind and a slow wind offers a worse fit ($\chi^2/\nu=1369.6/1430$) compared to the fast wind modelling. The ionization of the slow wind component is poorly constrained within the allowed parameter range. 
A very high relativistic outflowing velocity of $v=0.166^{+0.008}_{-0.009}c$ is still required in addition to a fast wind component obtained in Section \ref{fast}. In this case, a highly super-solar Ca abundance ($Z^{\prime}_{\rm Ca}>70Z_{\odot}$) is required to fit the 4.6~keV line, given the lack of the other absorption lines. So although the fast wind model does require an unusual Ar abundance, the slower wind model requires an even more extreme abundance ratio. 

Third, we explored the possibility of a warm absorber solution for the 4.6~keV absorption feature by limiting the velocity of the \texttt{warmabs} absorber within $v$<1000\,km\,s$^{-1}$ and assuming solar abundances. Warm absorbers are partially ionzied, optically thin, circumnuclear materials causing a series of absorption features \citep[e.g.][]{mckernan07} and are not uncommn among nearby AGNs \citep{reynolds97,george98,laha14}. However, the additional \texttt{warmabs} model fails to improve the fit of the 4.6~keV absorption line. Because no strong absorption lines of other elements are produced around 4.6~keV at an ionization state of $\log(\xi)=0-4$ without any other absorption features being produced simultaneously. For example, if the 4.6~keV absorption line is interpreted as the Ca~\textsc{xix} absorption line at 4.58~keV at an ionization state of $\log(\xi)=2$, a stronger Ca~\textsc{xiv} absorption line at 3.76~keV with an optical depth 1000 times larger than Ca~\textsc{xix} is expected at the same ionization state but not detected in the spectra.

Fourth, another possible origin of the 4.6~keV line feature is the Fe~\textsc{xxv/xxvi} absorption from a relativistic inflow \citep[e.g. Mrk~509][]{dadina05}. An inflowing velocity of $0.38c$ and $0.35c$ is required if the absorption feature corresponds to the Fe~\textsc{xxv} and Fe~\textsc{xxvi} line respectively. However, absorption features from an inflow with such a high velocity ($>0.35c$) have not been confirmed in other sources. 

In conclusion, we argue that the fast wind model that can account for both the 4.6~keV and 10~keV absorption features simultaneously is preferred.

\subsection{Spectral Analysis Results}

According to the best-fit reflection model, the disk is truncated with an inner radius of $R_{\rm in}=15^{+8}_{-5} r_{g}$. The disk viewing angle is well constrained at $i=23^{+3}_{-2}$$^{\circ}$ with respect to the normal direction of the disk. An additional distant reflection model \texttt{xillver} with the abundance linked to the disk reflection component only improves the fit by $\Delta\chi^2=4$ and gives a disk viewing angle of $i=24^{+4}_{-5}$$^{\circ}$.   

\begin{figure}
\centering
\includegraphics[width=\hsize]{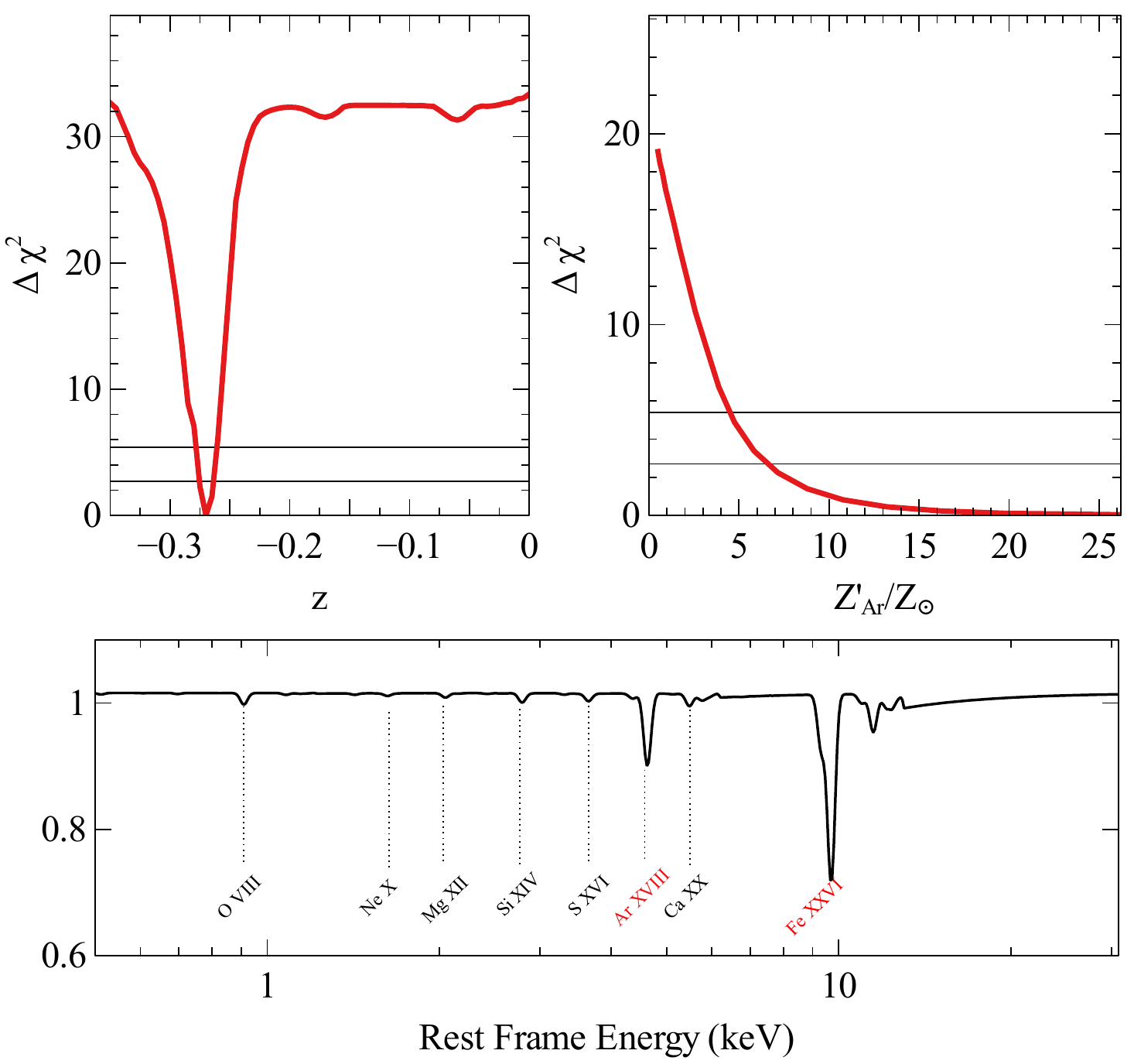}
\caption{Top left: the constraint on the redshift parameter $z$ of the absorber model \texttt{xstar}. Top right: the constraint on the Ar abundance $Z^{\prime}_{\rm Ar}$ of the absorber when being fitted as a free independent parameter as well as $Z^{\prime}_{\rm Fe}$. Bottom: The best-fit \texttt{xstar} model applied to a simple powerlaw continuum. The detected lines are marked in red.}
\label{pic_contour}
\end{figure}

The absorber model \texttt{xstar} prefers the Fe~\textsc{xxvi} interpretation of the broad absorption line at 10~keV, as it can fit the absorption lines at 4.6~keV and 10~keV simultaneously (see the left panel of Fig.\ref{pic_contour}). The outflow has a column density of $N^{\prime}_{\rm H} = 12^{+9}_{-4}\times10^{22}$\,cm$^{-3}$, an ionization of $\log(\xi^{\prime}$/erg\,cm\,s$^{-1})$ = $3.9^{+0.4}_{-0.3}$ and a redshift of $z=-0.270\pm0.006$, corresponding to a line-of-sight outflowing velocity $v=-0.319^{+0.005}_{-0.008}c$ accounting for the relativistic corrections and the cosmological redshift of WKK~4438. When the UFO argon abundance $Z^\prime_{\rm Ar}$ and iron abundance $Z^\prime_{\rm Fe}$ are treated as independent free parameters, a lower limit on the argon abundance is obtained ($Z^\prime_{\rm Ar}>6$) and a slight iron over-abundance is required compared to solar ($Z^\prime_{\rm Fe}=2.6^{+1.9}_{-2.0}$). Note that the abundances are probably degenerate with the column density. More discussion about the uncommon argon abundance can be found in Section \ref{discussion}. 

\begin{table}
\caption{The best-fit model parameters. The model is \texttt{tbnew*xstar*relxilllp}. The \texttt{tbnew} model is used to account for the Galactic absorption. The 3--10~keV band flux of the \suzaku observation is  $(9.01\pm0.03)\times10^{-12}$erg\,cm$^{-2}$\,s$^{-1}$. The same band flux of the \nustar observation is  $(5.15\pm0.08)\times10^{-12}$erg\,cm$^{-2}$\,s$^{-1}$.}
\begin{tabular}{c|cc|c}
\hline\hline
Parameter & Unit & \suzaku & \nustar \\
\hline
$N_{\rm H}$ & $10^{22}$~cm$^{-2}$ & \multicolumn{2}{c}{0.529$^{+0.014}_{-0.027}$}  \\
\hline
$N^{\prime}_{\rm H}$ & $10^{22}$~cm$^{-2}$ & \multicolumn{2}{c}{$12^{+9}_{-4}$}  \\
$\log(\xi^{\prime})$& log(erg\,cm\,s$^{-1}$) & \multicolumn{2}{c}{$3.9^{+0.4}_{-0.3}$} \\
$Z^{\prime}_{\rm Fe}$ & $Z_{\odot}$ & \multicolumn{2}{c}{$2.6^{+1.9}_{-2.0}$} \\
$Z^{\prime}_{\rm Ar}$ & $Z_{\odot}$ & \multicolumn{2}{c}{>6} \\
Redshift $z$ & & \multicolumn{2}{c}{$-0.270\pm0.006$} \\
\hline
$h$ & $r_g$ & <32 & <40 \\
$R_{\rm in}$ & $r_g$ & \multicolumn{2}{c}{$15^{+8}_{-5}$} \\
$i$ & deg & \multicolumn{2}{c}{$23^{+3}_{-2}$} \\
$Z_{\rm Fe}$ & $Z_{\odot}$ & \multicolumn{2}{c}{$0.9^{+0.7}_{-0.2}$} \\
$\log(\xi)$ & log(erg\,cm\,s$^{-1}$) & $0.96^{+0.19}_{-0.23}$ & <2.7 \\
$R_{\rm refl}$ & & $1.6\pm0.4$ & $1.8^{+1.6}_{-0.8}$ \\
$\Gamma$ & & $2.05^{+0.02}_{-0.04}$ & $1.93^{+0.17}_{-0.09}$ \\
\hline
$\chi^2/\nu$ & & \multicolumn{2}{c}{1363.0/1433} \\
\hline\hline
\end{tabular}
\label{tab_fit}
\end{table}

\section{Monte-Carlo Simulations} \label{mc_simulation}

In this section, we present our estimation on significance of the outflow detection in the current archival data based on Monte-Carlo simulations. 

We simulated 7000 sets of \suzaku (FI and BI) and \nustar (FPMA and FPMB) spectra using the best-fit \texttt{relxilllp} model discussed in Section \ref{spectral_analysis}. The same exposures and the flux levels as the real observations are considered. The key reflection model parameters used are the best-fit values obtained for the \texttt{relxilllp}-only model in Section \ref{spectral_fit} (h=5.06\,$r_g$, $i=23$$^{\circ}$, $R_{\rm in}=16r_{g}$, and $Z_{\rm Fe}=0.91$). The reflection fraction is $R_{\rm refl}=1.52$ for the \suzaku simulations and $R_{\rm refl}=2.38$ for \nustar simulations. The powerlaw continuum has a photon index of $\Gamma=2.05$ for \suzaku simulations and $\Gamma=1.99$ for \nustar simulations. The \texttt{fakeit} command is used to simulate the spectra. All the spectra are grouped to have a minimum count of 50 per bin using GRPPHA. We analysed each set of simulated spectra with both \texttt{relxilllp} and \texttt{xstar*relxilllp} models and measure the distribution of the statistical improvements achieved by adding the \texttt{xstar} absorber. All the parameters in Table \ref{tab_fit} are free to vary during our spectral fitting processes. We conducted a search for any absorption features in each set of simulated spectra by stepping the redshift parameter $z$ of the \texttt{xstar} model in the same way as in the real spectral analysis. 

The inclusion of the \texttt{xstar} model improves the fit statistic by $\Delta\chi^2=33.1$ in the real spectral analysis. The fit improvements in 17 out of the 7000 simulations exceed this threshold (a chance probability of 0.024), which means our detection of UFO in the current data is slightly higher than 3$\sigma$ significance by combining two observations. 

\section{Discussion} \label{discussion}

Archival \suzaku and \nustar data have revealed the NLS1 WKK~4438 shows evidence for both relativistic disk reflection and an ultra-fast outflow. The best-fit UFO parameters are a column density of $N^{\prime}_{\rm H}=12^{+9}_{-4}\times10^{22}$\,cm$^{-2}$, an ionization state of $\log(\xi^{\prime}$/erg\,cm\,s$^{-1})=3.9^{+0.4}_{-0.3}$, a slight iron over-abundance ($2.6^{+1.9}_{-2.0}Z_{\odot}$), an argon over-abundance (>6$Z_{\odot}$), and an outflowing velocity of $v_{\rm out}=0.319^{+0.005}_{-0.008}c$. The UFO absorption features are consistent when fitting the continuum only with the distant reflection model \texttt{xillver}. The inferred line significance is however higher with a distant reflection modelling. We discuss the physics properties of the outflow, other systematic uncertainties of the measurements, and future work in this section.

To calculate the disk accretion rate, we apply an average bolometric correction factor $\kappa=20$ \citep{bolometric} to the 2--10\,keV flux in the \suzaku and \nustar observation. The bolometric luminosity of WKK~4438 is estimated to be $L_{\rm bol}=1.6-2.8\times10^{44}$\,erg\,s$^{-1}=0.4-0.7L_{\rm Edd}$, assuming $M_{\rm BH}=2\times10^6$\,$M_{\odot}$ \citep{malizia08}. A combination of high accretion rate and UFO is seen in other sources, such as rapidly accreting AGNs with high accretion rate (e.g. IRAS~13224$-$3809: \cite{parker17a,parker17b,pinto17tmp}; 1H0707$-$495: \cite{dauser12}; PDS~456 \cite{nardini15,reeves18_2}; PG~1211$+$143: \cite{fukumura15,reeves18}), and ULXs \citep{pinto16,walton16,kosec18}. The Eddington accretion rate for a black hole of $M_{\rm BH}=2\times10^6$\,$M_{\odot}$ is $9\times10^{20}$\,kg\,s$^{-1}$. Therefore the mass accretion rate of WKK~4438 is $\dot {M}_{\rm in}\approx0.006-0.01M_{\odot}$\,yr$^{-1}$.

We estimate the mass outflow rate by following \citet{tombesi17}. A lower limit on the location of the wind can be derived from $r=2GM_{\rm BH}/v_{\rm out}^2\approx5.8\times10^{10}$\,m, which means that the wind is launched at $r\geq 20r_g$ away from the central SMBH. The mass outflow rate of the wind is $\dot {M}_{\rm out}=4 \pi \mu m_{p} r N_{\rm H} C_{\rm F} v_{\rm out}$, where $\mu=1.4$ is the average atomic mass per proton, $C_{\rm F}$ is the covering factor of the wind, and $m_{p}$ is the proton mass. Therefore, the lower limit on the mass outflow rate is $\dot {M}_{\rm out}\approx0.003C_{\rm F}M_{\odot}$yr$^{-1}$. The covering factor $C_{\rm F}$ of the wind remains unknown. The mass outflow rate is comparable with the mass accretion rate, assuming a maximum covering factor ($C_{\rm F}\approx1$ for PDS~456 \citep{nardini15} and IRAS~F11119$+$3257 \citep{tombesi15}). The kinetic power of the wind is then $\dot E_{\rm wind}=(1/2)\dot {M}_{\rm out}v_{\rm out}^2\approx9.59\times10^{42}$~erg\,s$^{-1}\approx3-5\%L_{\rm bol}$, assuming maximum covering factor. The low $\dot E_{\rm wind}/L_{\rm bol}$ ratio is due to a small column density of the wind and indicates the energetics might not be high enough to have an influence in AGN feedback \citep[e.g.][]{dimatteo05}. However, note that both the inferred velocity and the column density could be lower limits owing to the low inclination we infer, assuming the wind still has a roughly equatorial geometry.

In order to understand the high argon abundance required by the best-fit spectral model, we first investigated relative argon abundance against several other elements, such as silicon, sulfur, calcium, iron and oxygen, by fitting the spectra with the \texttt{warmabs} model. It turns out that the spectral fitting requires a high $Z^{\prime}_{\rm{Ar}}/Z^{\prime}_{\rm{Fe}}>4.2$ in the outflow. When the UFO and disk iron abundances are linked during the fit ($Z^\prime_{\rm Fe}=Z_{\rm Fe}$), a solar iron abundance is required ($Z^{\prime}_{\rm Fe}=Z_{\rm Fe}=0.96^{+0.66}_{-0.18}Z_{\odot}$) with $\chi^2=1364.1$, slightly worse than the fit with $Z^{\prime}_{\rm Fe}$ and $Z_{\rm Fe}$ both as a free parameter. The other parameter values do not change too much when the two iron abundances are linked. Second, we also tried fitting the data with a single matallicity ($Z^{\prime}$), with the abundance of all elements heavier than He linked together. It offers a metallicity value of $Z^{\prime}>0.6Z_{\odot}$ with $\chi^2=1388.4$. Finally, if we allow for a free iron abundance in addition to a single metallicity for the rest of the heavy elements, we obtained a fit with $\chi^2=1381.2$ ($Z^\prime_{\rm Fe}<3Z_{\odot}$ and $Z^{\prime}>0.6Z_{\odot}$) and only the 10~keV absorption feature being fitted. 

We note that \citet{tombesi10} report a similar scenario, where absorption from iron and argon with a common blueshift, and no other lines detected, for the Seyfert galaxy NGC~7582. \citet{dauser12} analysed the 1H0707-495 \xmm spectra and found a possible P Cygni profile of H-like Ar in the 2008 observations, where the redshift of the Ar line feature is however different from other absorption lines.  In contrast, another NLS1 IRAS~13224$-$3809 shows a series of absorption lines in the middle energy band, including Ne~\textsc{x}, S~\textsc{xvi} and Si~\textsc{xiv} absorption lines, but shows no evidence of Ar~\textsc{xviii} absorption line in the spectra \citep{jiang18_tmp}. Although, the variability spectrum of IRAS~13224$-$3809 shows some evidence of Ar~\textsc{xviii} feature \citep{parker17b}.

By fitting the spectra with the relativistic reflection model \texttt{relxilllp}, we obtain a disk viewing angle of $i=23^{+3}_{-2}$$^{\circ}$. Such a small viewing angle is unusual for a source with visible UFO absorption features due to the opening angle of the wind. Other sources where UFO is detected show evidence of large disk viewing angle when being fitted with relativistic disk reflection model, such as 1H0707$-$495 \citep[$i\approx50$$^{\circ}$,][]{dauser12}, PDS~456 \citep[$i=65\pm2$$^{\circ}$,][]{chiang17}, IRAS~13224$-$3809 \citep[$i=67\pm3$$^{\circ}$,][]{jiang18_tmp}. An alternative explanation of the absorption features is ionized materials corotating above the disk, where the relativistic velocities occur naturally \citep{gallo11}. Such a model has been successfully applied to PG~1211$+$143 \citep{gallo13} and IRAS~13224$-$3809 (Fabian et al. in prep). However, this scenario seems to be unlikely for WKK~4438 due to its small viewing angle and truncated disk. 

It is now becoming apparent that these extreme outflows are variable phenomena in various sources, such as IRAS~13224$-$3809 \citep{parker17a,pinto17tmp}, Mrk~509 \citep{cappi09}, PDS~456 \citep{reeves09,nardini15,matzeu1}, and PG1211$+$143 \citep{pounds03,reeves18}. In many of these cases, the black hole masses are high (e.g. $M_{\rm BH}\approx10^9$M$_{\odot}$ for PDS~456 and $M_{\rm BH}\approx10^8$M$_{\odot}$ for PG1211$+$143). However, similar to IRAS~13224$-$3809, the mass of the black hole in WKK~4438 is rather low ($M_{\rm BH}\approx2\times10^6$\,$M_{\odot}$, \citet{malizia08}). This means the outflow in WKK~4438 is potentially of particular interest, as variability timescales are generally expected to scale with black hole mass. It may therefore be possible to study the variability of the outflow in WKK~4438 -- and any potential response to intrinsic changes in the source -- with a few deep observations in the future. Unfortunately, the data currently available do not have sufficient signal-to-noise or broadband coverage to undertake more studies at the present time.

\section*{Acknowledgements}

J.J. acknowledges support by the Cambridge Trust and the Chinese Scholarship Council Joint Scholarship Programme (201604100032). D.J.W. acknowledges support from an STFC Ernest Rutherford fellowship. M.L.P. is supported by a European Space Agency (ESA) Research Fellowship. A.C.F. acknowledges support by the ERC Advanced Grant 340442. This work made use of data from the \nustar mission, a project led by the California Institute of Technology, managed by the Jet Propulsion Laboratory, and funded by NASA, and data obtained from the Suzaku satellite, a collaborative mission between the space agencies of Japan (JAXA) and the USA (NASA). This research has made use of the \nustar Data Analysis Software (NuSTARDAS) jointly developed by the ASI Science Data Center and the California Institute of Technology. 

%%%%%%%%%%%%%%%%%%%%%%%%%%%%%%%%%%%%%%%%%%%%%%%%%%



%\bibliographystyle{mnras}
%\bibliography{wkk4438.bib} % if your bibtex file is called example.bib

%%%%%%%%%%%%%%%%%%%%%%%%%%%%%%%%%%%%%%%%%%%%%%%%%%

%%%%%%%%%%%%%%%%% APPENDICES %%%%%%%%%%%%%%%%%%%%%

%\appendix



%%%%%%%%%%%%%%%%%%%%%%%%%%%%%%%%%%%%%%%%%%%%%%%%%%

% Don't change these lines
\bsp	% typesetting comment
\label{lastpage}
\end{document}